\documentclass[fleqn,usenatbib,useAMS]{mnras}
 \usepackage{graphicx}
 \usepackage{amsmath}
 \usepackage{amssymb}
 \usepackage[T1]{fontenc}
 \usepackage{ae,aecompl}
 \usepackage{txfonts}

 \title[Dynamical groupings among NEOs]
       {Far from random: dynamical groupings among the NEO population}

 \author[C. de la Fuente Marcos and R. de la Fuente Marcos]
        {C.~de~la~Fuente~Marcos\thanks{E-mail: carlosdlfmarcos@gmail.com}
         and
         R. de la Fuente Marcos \\
         Apartado de Correos 3413, E-28080 Madrid, Spain}
 \date{Accepted 2015 December 4.
       Received 2015 December 3;
       in original form 2015 September 7}
 \pubyear{2016}
 
\begin{document}
  \label{firstpage}
  \pagerange{\pageref{firstpage}--\pageref{lastpage}}
  \maketitle

  \begin{abstract}
     Among the near-Earth object (NEO) population there are comets and 
     active asteroids which are sources of fragments that initially move 
     together; in addition, some NEOs follow orbits temporarily trapped in 
     a web of secular resonances. These facts contribute to increasing the 
     risk of meteoroid strikes on Earth, making its proper quantification 
     difficult. The identification and subsequent study of groups of small 
     NEOs that appear to move in similar trajectories are necessary steps 
     in improving our understanding of the impact risk associated with  
     meteoroids. Here, we present results of a search for statistically 
     significant dynamical groupings among the NEO population. Our Monte 
     Carlo-based methodology recovers well-documented groupings like the 
     Taurid Complex or the one resulting from the split comet 
     73P/Schwassmann-Wachmann 3, and new ones that may have been the source 
     of past impacts. Among the most conspicuous are the Mjolnir and Ptah 
     groups, perhaps the source of recent impact events like Almahata Sitta 
     and Chelyabinsk, respectively. Meteoroid 2014~AA, that hit the Earth 
     on 2014~January~2, could have its origin in a marginally significant 
     grouping associated with Bennu. We find that most of the substructure 
     present within the orbital domain of the NEOs is of resonant nature, 
     probably induced by secular resonances and the Kozai mechanism that 
     confine these objects into specific paths with well-defined perihelia. 
  \end{abstract}


  \section{Introduction}
     It is a well-known fact that, as a result of observational bias and planetary perturbations, the distribution of the orbital elements 
     of the near-Earth object (NEO) population is significantly non-random (see e.g. JeongAhn \& Malhotra 2014). The largest known 
     dynamically quasi-coherent subpopulation is that of the so-called Taurid Complex asteroids (Steel, Asher \& Clube 1991) with two groups 
     of objects with apsidal lines approximately aligned with those of asteroid 2212 Hephaistos (1978 SB) and comet 2P/Encke (Asher, Clube 
     \& Steel 1993). This stream of material could be made of debris left behind by a large comet that disintegrated tens of thousands of 
     years ago (Steel et al. 1991), but dynamically resonant substructure is also possible (Soja et al. 2011; Sekhar \& Asher 2013). Another 
     obvious source of dynamical coherence among NEOs can be traced back to comets and active asteroids (Jewitt 2012) that produce fragments 
     which initially move in unison. 

     Although dynamical coherence among NEOs should exist, no firm evidence on this matter has been produced yet. Schunov\'a et al. (2012) 
     have shown that robust statistical estimates of a dynamical link between NEOs require groups of four or more objects, but they could 
     not find any statistically significant group of dynamically related NEOs among 7\,563 objects. However, in Schunov\'a et al. (2012) the
     significance of asteroid clusters was computed by comparing them to a scrambled version of the observed distribution in order to have a 
     realistic but cluster-free background to compare against, and they used the $D$-criterion of Southworth \& Hawkins (1963, $D_{\rm SH}$) 
     as metric. The presence of dynamical coherence among NEOs is an interesting topic by itself, but it is also of considerable importance 
     to the subject of planetary defence as it makes the quantification of the risk of meteoroid strikes on Earth difficult. Here, we 
     compute statistical significance maps for the currently known NEO population and study the possible presence of robust dynamical 
     groupings. This paper is organized as follows. In Section 2, we discuss both the data and the methodology used in our analysis. 
     Significance maps are presented in Section 3. Our results are validated using multiple techniques in Section 4. Relevant dynamical 
     groupings are explored in Section 5. Section 6 discusses our results and their implications. A summary of our conclusions is given in 
     Section 7.

  \section{Data and methodology}
     In order to identify statistically significant dynamical groupings among the NEO population we use the list of NEOs currently 
     catalogued (as of 2015 November 5, 13\,392 objects) by the Jet Propulsion Laboratory's (JPL) Solar System Dynamics Group (SSDG) 
     Small-Body Database (SBDB),\footnote{http://ssd.jpl.nasa.gov/sbdb.cgi} and the $D$-criteria of Lindblad \& Southworth (1971), $D_{\rm 
     LS}$, and that of Valsecchi, Jopek \& Froeschl\'e (1999), $D_{\rm R}$. Therefore, we focus on semimajor axis, $a$, eccentricity, $e$, 
     and inclination, $i$, for objects with perihelion distance, $q=a(1-e)$, $q<1.3$~au. Given the statistical nature of our methodology, 
     the fact that the information in the data base is incomplete and biased is irrelevant as we seek statistically significant dynamical 
     groupings among known NEOs. Most of these objects have been observed because they passed close to our planet. If they are close passers, 
     this sample corresponds to bodies that may eventually collide with the Earth. In fact, some of them have small but non-zero impact 
     probability. The list includes 13\,220 near-Earth asteroids (NEAs) and 172 near-Earth comets (NECs). 

\begin{figure}
        \centering
         \includegraphics[width=\linewidth]{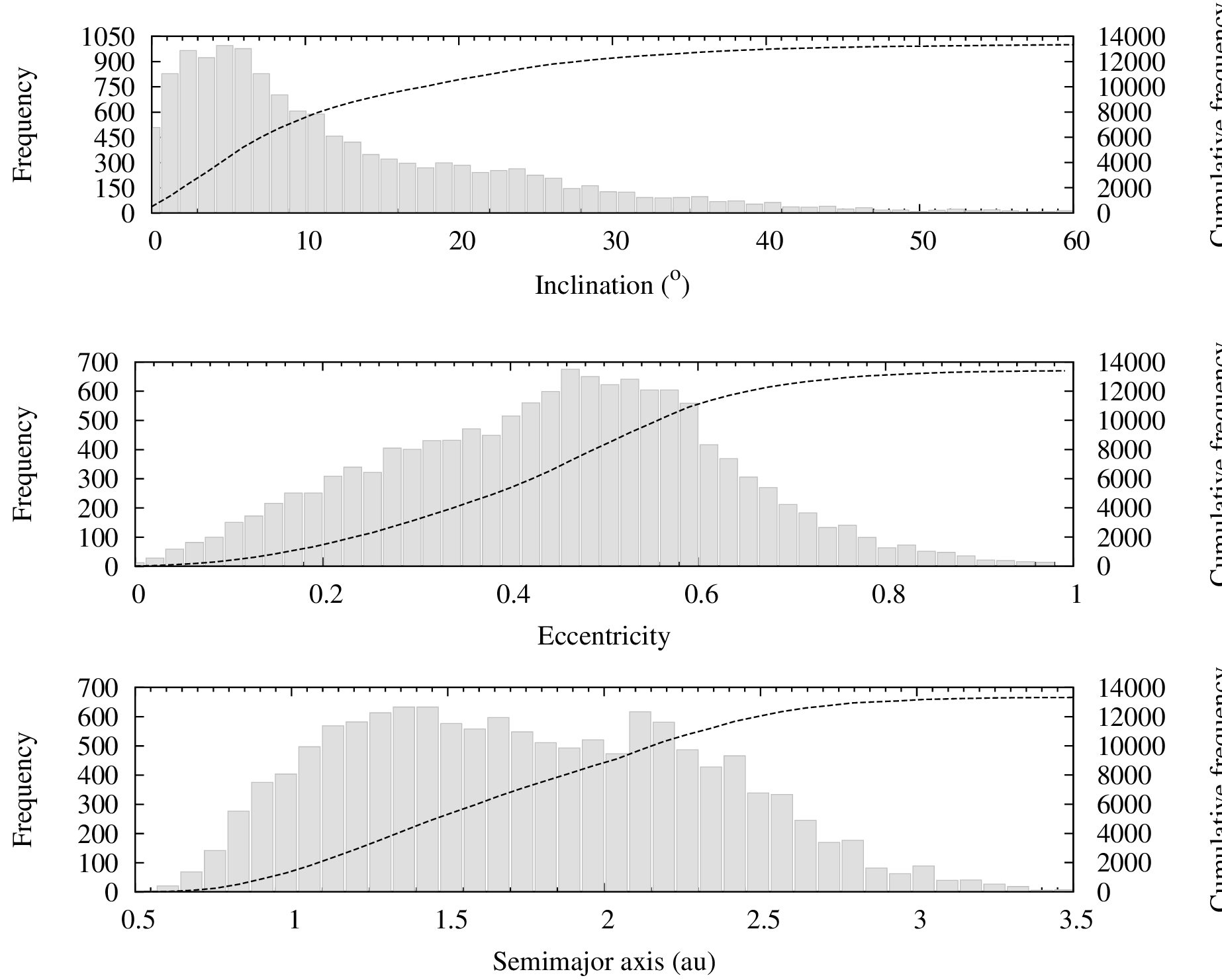}
         \caption{Distribution of the values of the orbital elements $a$, $e$, and $i$ for the sample of NEOs used in this study (see the 
                  text for details). Here and in subsequent histograms, the bin width has been computed using the Freedman-Diaconis rule, 
                  i.e. $2\ {\rm IQR}\ n^{-1/3}$, where IQR is the interquartile range and $n$ is the number of data points which is 13\,392.
                  Data from the JPL's SSDG SBDB as of 2015 November 5, 13\,220 NEAs and 172 NECs.
                 }
         \label{histo}
      \end{figure}

     Figure \ref{histo} shows the distribution of the values of the orbital elements $a$, $e$, and $i$ for the sample used in this study. In 
     this and subsequent histograms, the bin width has been computed using the Freedman-Diaconis rule (Freedman \& Diaconis 1981), i.e. $2\ 
     {\rm IQR}\ n^{-1/3}$, where IQR is the interquartile range and $n$ is the number of data points. Most objects can be found in the 
     region within $1.0<a<2.5$~au, $0.3<e<0.6$, and $1<i<10$\degr. The mean values of $a$, $e$, and $i$ are 1.93 au, 0.46, and 13\fdg35, 
     respectively; the median values of $a$, $e$, and $i$ are 1.75 au, 0.47, and 9\fdg47, respectively. The respective IQR values are 0.90 
     au, 0.25, and 14\fdg12; the upper quartiles are 2.23 au, 0.58, and 19\fdg04, respectively. 
%
%
%
%

     We have generated a large number of virtual NEOs under the form of sets of orbital elements with $q<1.3$~au, $e\in(0, 0.99)$, $i\in(0, 
     60)$\degr, longitude of the ascending node, $\Omega\in(0, 360)$\degr, and argument of perihelion, $\omega\in(0, 360)$\degr. These 
     parameters are randomly sampled from a uniform distribution. For each set, we have computed the $D$-criteria with all the NEOs 
     currently catalogued and counted the number of minor bodies with both $D_{\rm LS}$ and $D_{\rm R} < 0.05$. These reference values of 
     $D$ are reasonable when associating meteors with their parent bodies (see e.g. Rudawska, Vaubaillon \& Atreya 2012). In summary, we 
     generate a random virtual NEO within a chosen volume of the parameter space and count how many real NEOs have both $D_{\rm LS}$ and 
     $D_{\rm R} < 0.05$ with respect to the virtual one. We repeat this step mutiple times in order to sample properly the chosen volume of 
     the parameter space, creating a three-dimensional random mesh. In this way, we compute the average number of real NEOs ---and its 
     standard deviation--- that satisfy the assumed constraints in $D$ for sets of orbital elements generated uniformly. Using these values, 
     we can readily estimate the statistical significance of any putative NEO clustering, $|c - \langle{c}\rangle|/\sigma_{\rm c}$, where 
     $c$ is the count number for a given set of elements and $\langle{c}\rangle$ and $\sigma_{\rm c}$ are, respectively, the average value 
     and its standard deviation for the entire numerical experiment. 

     In the following, we consider that any set of orbital elements with an associated number of bodies under 3$\sigma$ corresponds to an 
     assumed background population for which no dynamical coherence different from a random one is present. Sets of orbital elements with 
     associated number of bodies above 5$\sigma$ are singled out for further study. Given the discrete nature of our approach, sets of 
     orbital elements relatively close to each other in parameter space have comparatively high (or low) numbers of associated bodies. This 
     lets us produce statistical significance maps in terms of $\sigma$ and easily identify the regions of the studied volume in orbital 
     parameter space for which the number of known minor bodies is substantially above the average or, what we call, the background (see
     Section 4 for additional details). 

     It may be argued that many of the orbits available from SBDB are based on short or very short arcs and, therefore, this will make our
     conclusions less robust or even entirely questionable. Fortunately, given two sets of orbital elements, one fixed and the second one 
     variable within reasonable tolerances (i.e. $<20$ per cent) the resulting values of $D_{\rm LS}$ and $D_{\rm R}$ are still not too 
     different and in any case close enough to make any variations in the final count virtually irrelevant for any practical purposes. In 
     addition, the large number of trials performed further contributes to minimizing any possible harmful effects. As an internal 
     consistency test, the clusterings with the highest statistical significance ($\sim$13.5$\sigma$ with $\sim$70 members) are invariably 
     associated with the most recent documented fragmentation episode within the NEO orbital realm, that of comet 73P/Schwassmann-Wachmann 3 
     (see e.g. Weaver et al. 2006) with $a\sim3$ au, $e\sim0.7$, and $i\sim11$\degr.

  \section{Statistical significance maps}
\begin{figure*}
       \centering
        \includegraphics[width=0.49\linewidth]{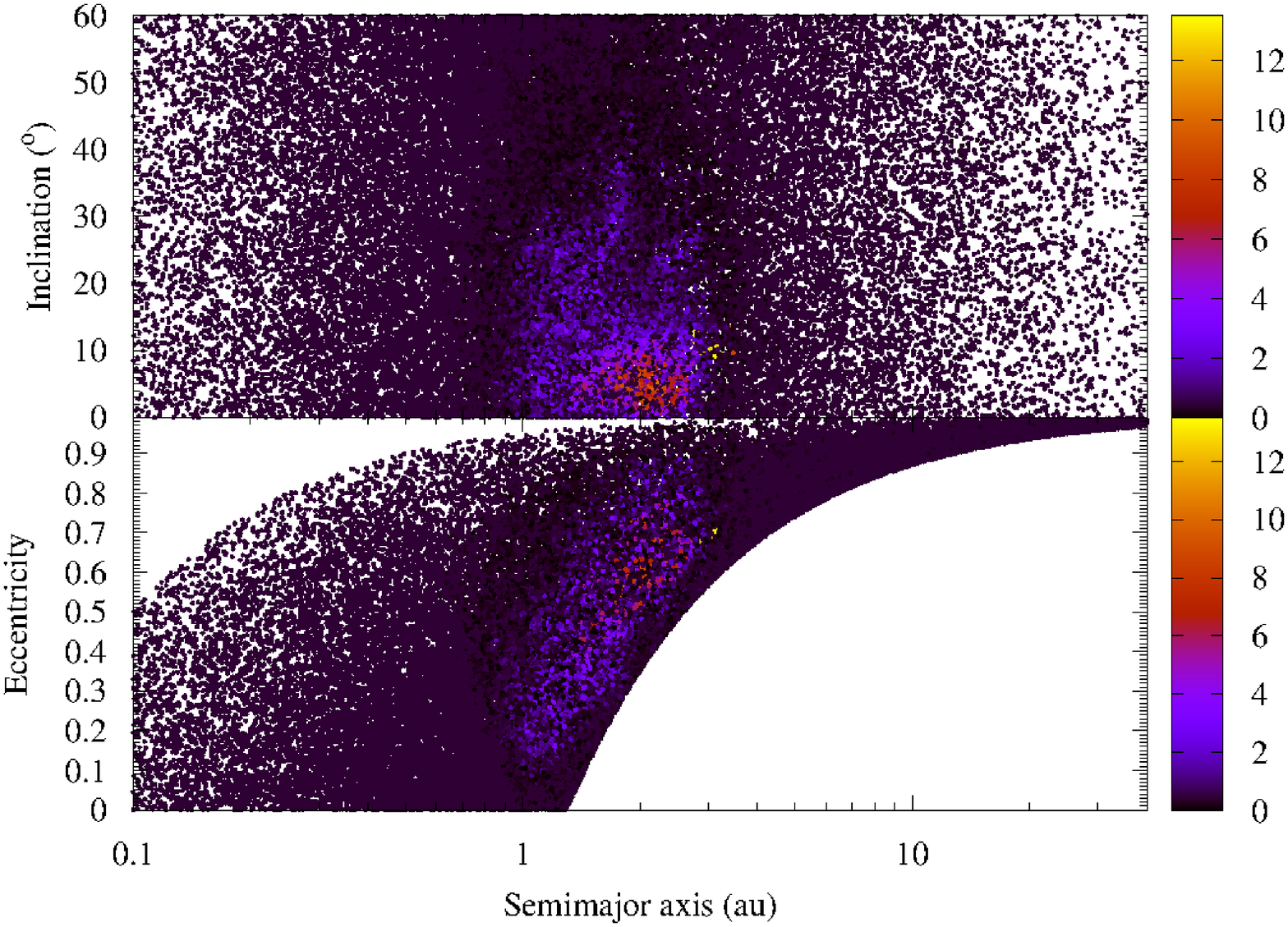}
        \includegraphics[width=0.49\linewidth]{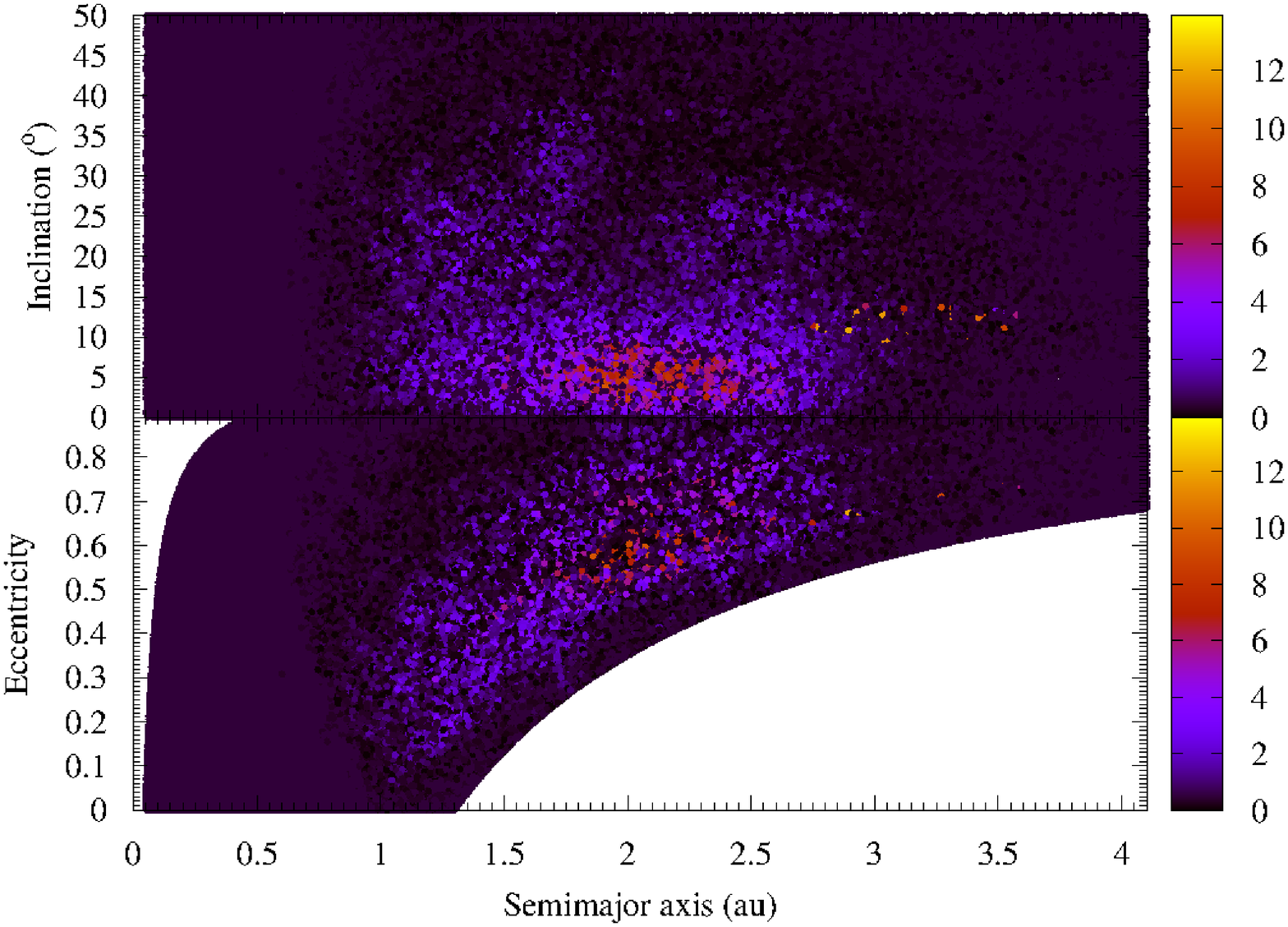}\\
        \caption{Results of our statistical analysis. The colours in the colour maps are proportional to the value of the statistical 
                 significance (see the text for details). The left-hand side panels correspond to 1$\times10^6$ sets of orbital elements
                 with $q<1.3$~au, $e\in(0, 0.99)$, $i\in(0, 60)$\degr, $\Omega\in(0, 360)$\degr, and $\omega\in(0, 360)$\degr; the 
                 right-hand panels show the results of another experiment of 5$\times10^5$ sets with $q<1.3$~au but $e\in(0, 0.9)$, $i\in(0, 
                 50)$\degr.
                }
        \label{maps}
     \end{figure*}

     Figure \ref{maps} summarizes our statistical analysis; the colours in the colour maps are proportional to the value of the statistical
     significance in units of $\sigma$. Two different experiments are shown: 1$\times10^6$ sets of orbital elements with $q<1.3$~au, 
     $e\in(0, 0.99)$, $i\in(0, 60)$\degr, $\Omega\in(0, 360)$\degr, and $\omega\in(0, 360)$\degr (left-hand panels) and 5$\times10^5$ sets 
     with $q<1.3$~au but $e\in(0, 0.9)$ and $i\in(0, 50)$\degr (right-hand panels). For the first statistical experiment 
     $\langle{c}\rangle=2.17\pm4.96$ and for the second one $\langle{c}\rangle=2.78\pm5.54$ NEOs; the errors quoted correspond to the values 
     of the standard deviation. In both experiments we use the same input data, the 13\,392 NEOs, but different resolutions ---1$\times10^6$ 
     and 5$\times10^5$, respectively---, i.e. the size of the sample is always the same. For each experiment, we consider that those sets of 
     orbital elements with associated counts under 3$\sigma$ (see above and Section 4) are compatible with statistical fluctuations. On the 
     other hand, those with associated counts above 5$\sigma$ represent bona fide dynamical groupings (i.e. groups resulting from the 
     perturbing action of resonances) and, in some cases, perhaps even genetically related asteroid clusters. Each statistically significant 
     dynamical grouping is likely to include unrelated interlopers (compatible with the background population and perhaps as high as 10 per 
     cent in fraction) and statistically significant neighbouring groups may correspond to a single piece of coherent dynamical substructure 
     as the orbital subdomains overlap and our approach is discrete. Given the fact that spectroscopic information for individual asteroids 
     is scarce at present, it is not possible to investigate properly which fraction of clustered NEOs could be true, genetic siblings.
%
%
%
%

\begin{figure}
        \centering
         \includegraphics[width=\linewidth]{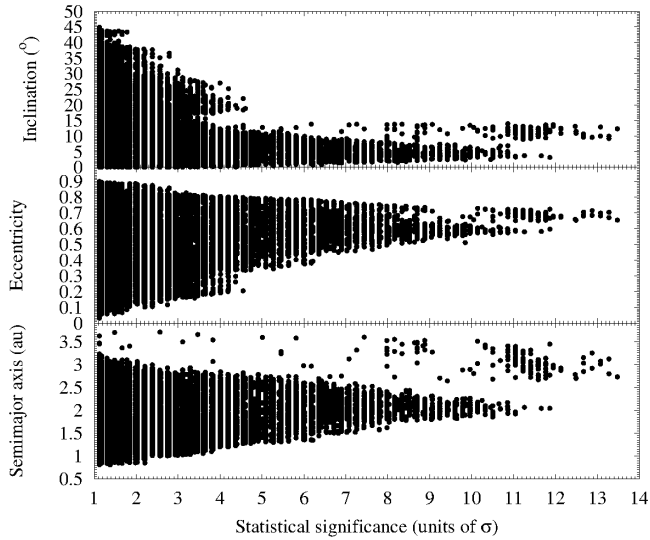}
         \caption{Counts with significance $> 1\sigma$. The apparent quantization of the data is induced by the discrete value of the 
                  counts and the underlying substructure.
                 }
         \label{all}
      \end{figure}

     In principle, the most striking feature in Fig. \ref{maps} is obviously the vast Taurid Complex. Steel \& Asher (1996) characterized 
     the orbital domain occupied by the Taurid Complex asteroids as $1.8<a<2.6$~au, $0.64<e<0.85$, and $i\leq12$\degr. Our significance maps 
     suggest a larger size although some of the statistically significant substructure is well outside the region outlined by Steel \& Asher 
     (1996), particularly if the value of $e$ is considered. In order to better understand the maps, we show in Fig. \ref{all} the values of 
     the orbital elements $a$, $e$, and $i$ as a function of the statistical significance of the groupings. Some regularities are clearly 
     seen, but also the presence of outliers. Not considering outliers, the best dynamical groupings have $a\sim2$~au, $e\sim0.6$, and 
     $i\sim4$\degr (with a significance of $\sim$12$\sigma$ and $\sim$60 members). The value of the eccentricity is at the edge of the most 
     probable range in Fig. \ref{histo}, and far from the median value of 0.47 and within the fourth quartile; the value of the inclination 
     is far from the median value of 9\fdg47 and belongs to the first quartile. These statistical facts suggest that any clustering found is 
     real and not linked to the most probable values of the orbital parameters of the sample studied. In other words, the observed dynamical 
     substructure is unlikely to be the result of harmful selection effects. On the other hand, the most significant dynamical groupings 
     (not outliers) are near the edge or outside the Taurid Complex as defined by Steel \& Asher (1996) which indicates that the majority of 
     the observed dynamical groupings may not be directly related to the Taurid Complex although it is possible that non-gravitational 
     forces may have populated orbits outside the Taurid Complex after the hierarchical disintegration of a giant comet as envisioned by 
     Steel et al. (1991). 
%
%
%
%

\begin{figure}
        \centering
         \includegraphics[width=\linewidth]{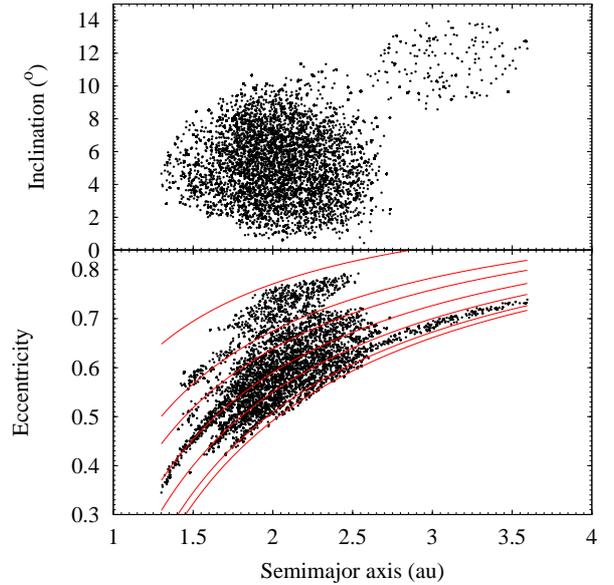}
         \caption{Dynamical groupings with statistical significance above 5$\sigma$. These dynamical groupings are likely to include objects 
                  which are dynamically or even genetically linked. The curves indicate the set of orbits having perihelion at (from top to
                  bottom): 0.458 au (Mercury's aphelion), 0.65 au, 0.72 au (Venus' semimajor axis), 0.82 au, 0.9 au, 0.983 au (Earth's 
                  perihelion), and 1.017 au (Earth's aphelion). 
                 }
         \label{best}
      \end{figure}

     If we focus on the dynamical groupings with statistical significance above 5$\sigma$ (see Fig. \ref{best}), we observe that a number of 
     them appear organized along well-defined tracks, curves of constant perihelion distance. This type of substructure is normally 
     associated with resonances. Mean-motion resonances in this region are relatively weak, but secular resonances where the precession of
     the node of the perihelion of a NEO relative to a planet librates could be strong. The effect of secular resonances in the region of 
     semimajor axes smaller than 2~au was first studied in a seminal paper by Michel \& Froeschl\'e (1997) and further explored in Michel 
     (1997, 1998). Michel \& Froeschl\'e (1997) found that objects with $0.4<a<0.65$~au and $0.82<a<0.9$~au are affected by secular 
     resonances (in particular with the Earth, Mars and Saturn) and those with $0.65<a<0.82$~au and $0.9<a<1.1$~au are affected by the 
     Kozai mechanism (Kozai 1962) that, at low inclination, induces libration of the argument of perihelion around 0\degr or 180\degr (see 
     e.g. Michel \& Thomas 1996). An argument of perihelion librating around 0\degr means that the orbit reaches perihelion at 
     approximately the same time it crosses the Ecliptic from South to North (the ascending node); a libration around 180\degr implies that 
     the perihelion is close to the descending node. Although their calculations were completed making a number of assumptions which are in 
     conflict with the range of parameters studied here (in particular, they focused on relatively low values of the eccentricity, 
     $e\sim0.1$), their findings are the key to understanding why the clustering tends to appear along tracks instead of in the form of 
     spots of more or less circular symmetry that may signal relatively recent asteroidal (or cometary) disruption events. The curves in Fig. 
     \ref{best} indicate the set of orbits having perihelion at the characteristic distances found by Michel \& Froeschl\'e (1997). The 
     curve corresponding to a value of the perihelion distance of 0.82 au reproduces one of the tracks quite well, strongly suggesting that 
     secular resonances shepherd asteroids around, sculpting the orbital architecture of the NEO population. Objects confined near the 
     curve of perihelion distance equal to 0.82~au reach perihelion within the region that separates the orbital domain where the Kozai 
     mechanism is dominant from that where secular perturbations are active (see figs 2 and 3 in Michel \& Froeschl\'e 1997). Similar 
     reasonings can be made for other substructures.

     Figure \ref{best} can be interpreted from an evolutionary point of view. Assuming that no major observational biases are at work (see 
     below), the areas of the figure where more objects are observed can be seen as intrinsically more stable and therefore more suitable to 
     retain NEOs in their neighbourhood for longer time-scales. Those devoid of objects are either dynamically unstable or prone to be 
     associated with impact orbits. In other words, NEOs following more stable orbits are more likely to be observed and those populating 
     sets of trajectories that intersect the orbits of the inner planets will be depleted more efficiently via actual impacts or dynamical 
     ejections following close encounters. This evolutionary interpretation has been previously used in Froeschl\'e et al. (1995) to 
     distinguish between fast tracks affected by strong and rapid changes in $e$ due to resonances and slow tracks characterized by a random 
     walk in $a$ due to close approaches to an inner planet. Unfortunately, those NEOs that pass closer to the Earth, even if they are very 
     small, are also the ones more likely to be detected no matter how unstable their orbits are. In summary, a fully evolutionary 
     interpretation is not possible in this case due to strong selection effects.

%
%
%
%

\begin{figure}
        \centering
         \includegraphics[width=\linewidth]{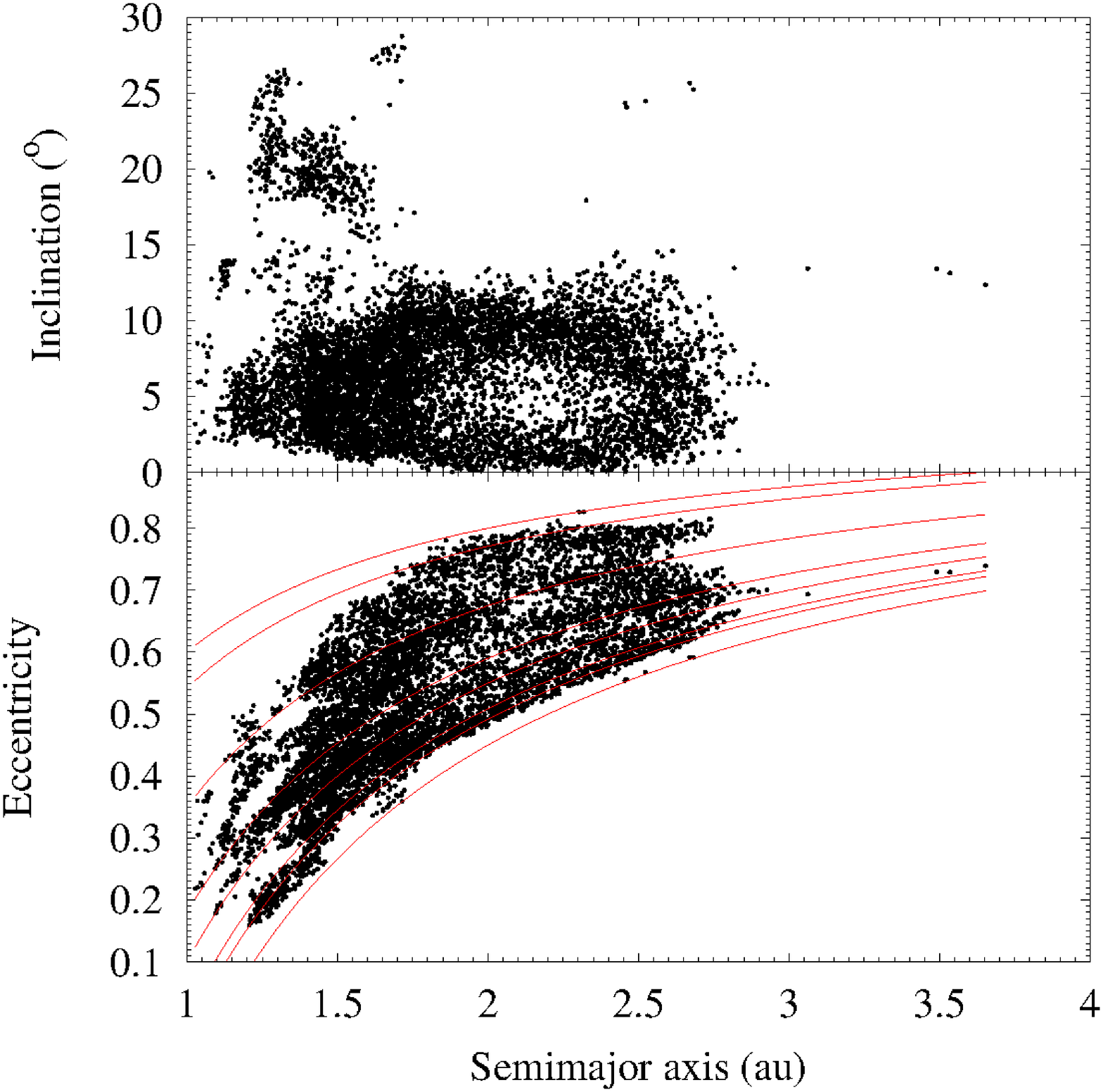}
         \caption{Same as Fig. \ref{best} but for groupings with statistical significance in the range 3--5$\sigma$. Most of these 
                  groupings are probably chance alignments. The curves indicate the set of orbits having perihelion at (from top to bottom): 
                  0.4 au, 0.458 au (Mercury's aphelion), 0.65 au, 0.82 au, 0.9 au, 0.983 au (Earth's perihelion), 1.017 au (Earth's 
                  aphelion), and 1.1 au.
                 }
         \label{soso}
      \end{figure}

      If we shift our focus to the marginally significant groupings with 3--5$\sigma$ (these groupings usually have 18 to 30 members), 
      further substructure is uncovered, including dynamical groupings with higher inclination and lower eccentricity. The level of 
      dynamical complexity is similar to the one found within the main asteroid belt (see e.g. Nesvorn\'y, Bro\v{z} \& Carruba 2015) 
      although the values of the eccentricity are higher. Figure \ref{soso} shows that relevant clustering is still mostly organized along 
      lines of constant perihelia although the evidence is not as clear as in the case of high-significance dynamical groupings (above 
      5$\sigma$) hinting at the presence of strong background contamination.
%
%
%
%

    In general, performance of statistical significance maps is affected by tessellation geometry (variation of cell size and cell shape) as 
    well as aggregation level (average cell size). They depend on counts on a grid cell and a single actual cluster that is bisected by a 
    large cell may appear either as not significant or as two distinct groupings depending on whether the large cell is included in the 
    studied zone. Our Monte Carlo-style (Metropolis \& Ulam 1949; Press et al. 2007) approach avoids these problems as there are no real 
    cells; our counts satisfy that both $D_{\rm LS}$ and $D_{\rm R} < 0.05$ and their centres are randomly chosen. As long as enough sets of 
    virtual orbital elements are examined, no statistically significant groupings can be missed.

  \section{Validation of the search algorithm}
     But, could it be that the statistically significant groupings found above are just statistical artefacts? In order to verify the 
     strength and reliability of our dynamical grouping recovery algorithm, we have tested our method on data from a uniform distribution, 
     the original data altered by adding Gaussian noise, and data from the synthetic NEOSSat-1.0 orbital model described in Greenstreet, Ngo 
     \& Gladman (2012). We have also studied the role of the value of the cut-off parameter and diverse techniques to subtracting the 
     background.

     \subsection{Characterizing a putative uniform background}
\begin{figure}
       \centering
        \includegraphics[width=\linewidth]{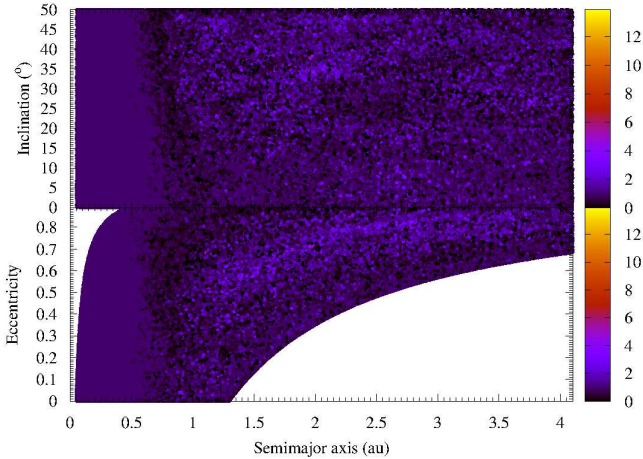}
        \caption{As Fig. \ref{maps}, right-hand panel but for synthetic data with uniformly distributed orbital elements (see the text for 
                 details). The data are uniform in $q$ but not in $a$.
                }
        \label{mapu}
     \end{figure}

        Let us assume that we have a population of NEOs with values of the orbital elements uniformly distributed within the domain 
        $q<1.3$~au (uniform in $q$ but not in $a$), $e\in(0, 0.9)$, $i\in(0, 50)$\degr, $\Omega\in(0, 360)$\degr, and $\omega\in(0, 
        360)$\degr. Figure \ref{histo} clearly shows that this assumption is incorrect, but it is still a useful approximation if we want to
        define a significance threshold on purely mathematical grounds. Such hypothetical and ideal population is not affected by resonances 
        and does not include any dynamical grouping nor genetic asteroid family. We have generated 13\,392 virtual objects following this 
        specification and subjected the resulting data to the same analysis described above, generating 5$\times10^5$ sets of orbital 
        elements also within the domain $q<1.3$~au, $e\in(0, 0.9)$, $i\in(0, 50)$\degr, $\Omega\in(0, 360)$\degr, and $\omega\in(0, 
        360)$\degr. The probability of having no dynamically related objects, $P(0)$, is 0.349632; $P(>21)=0$. For this experiment 
        $\langle{c}\rangle=3.03\pm3.11$ (see Fig. \ref{mapu}). In other words, a cluster of 22 virtual objects would represent a $6\sigma$ 
        deviation from the average value of $\sim$3 that characterizes a uniform background. This result supports our choice of 5$\sigma$ as 
        the suitable level above which statistically significant clustering could be present. Our statistically best groupings have all more 
        than 30 members (see above). Within the 5$\times10^5$ sets, the one with the highest statistical significance had 21 members which 
        implies that the sampling resolution (5$\times10^5$ sets) was high enough to identify correctly those groupings which are likely due 
        to statistical fluctuations. 
%
%
%
%

     \subsection{Gaussian noise}
        As an additional safety check, we have performed two numerical experiments in which we generated 13\,392 virtual objects using the 
        original data with Gaussian noise added. In these two experiments, the level of Gaussian noise is equivalent to 30 per cent and 90 
        per cent of the actual values of the parameters, respectively. In other words and for the first of these experiments, given a value 
        of an orbital parameter we add $r$ times 30 per cent of the value with $r$ being a Gaussian random number resulting from the 
        application of the Box-Muller method (Press et al. 2007). Adding noise simulates the role played by errors in orbit determination. 
        These scrambled-by-noise realizations can also be used as realistic but cluster-free backgrounds to compare against, particularly
        the one generated by adding a level of Gaussian noise as high as 90 per cent (see below). Figure \ref{map30} is equivalent to the 
        right-hand panel in Fig. \ref{maps} after adding Gaussian noise at the 30 per cent level. For this experiment 
        $\langle{c}\rangle=2.31\pm4.50$, and the clustering with the highest significance (9.5$\sigma$) includes 45 members and has $a=1.6$ 
        au, $e=0.63$, and $i=5\fdg1$. Figure \ref{map90} shows the results of adding Gaussian noise at the 90 per cent level. For this new 
        experiment $\langle{c}\rangle=0.77\pm1.55$, and the clustering with the highest significance (13.7$\sigma$) includes 22 members and 
        has $a=2.1$ au, $e=0.87$, and $i=3\fdg0$. Therefore, we find that the statistically significant dynamical groupings found above are 
        statistically robust as they disappear for increasingly randomized data.
%
%
     \begin{figure}
       \centering
        \includegraphics[width=\linewidth]{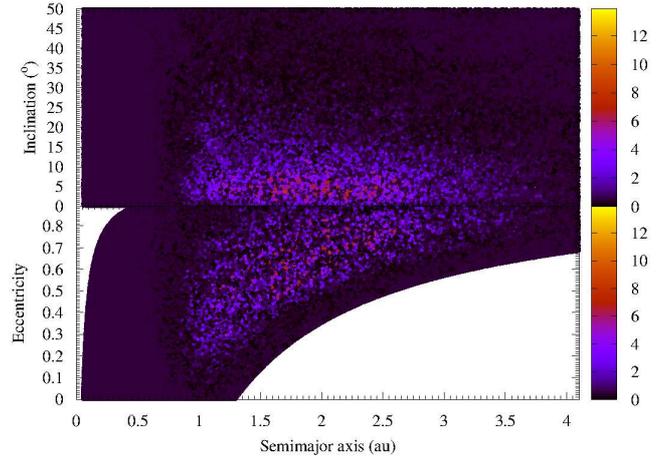}
        \caption{As Fig. \ref{maps}, right-hand panel but for altered data with Gaussian noise at the 30 per cent level (see the text for 
                 details).
                }
        \label{map30}
     \end{figure}
%
%
%
%
     \begin{figure}
       \centering
        \includegraphics[width=\linewidth]{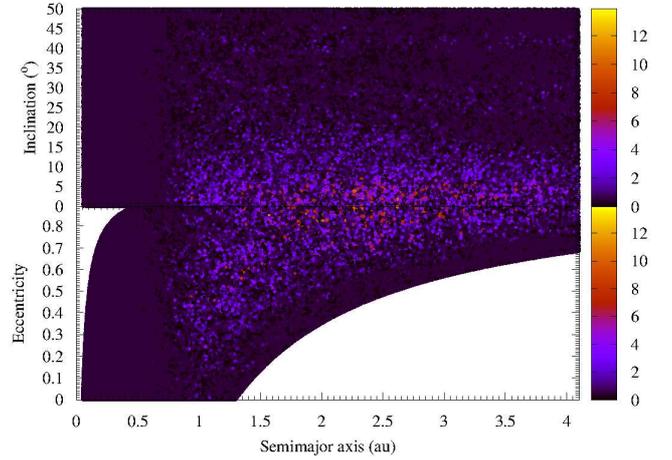}
        \caption{As Fig. \ref{maps}, right-hand panel but for altered data with Gaussian noise at the 90 per cent level (see the text for 
                 details).
                }
        \label{map90}
     \end{figure}
%
%

     \subsection{NEOSSat-1.0 orbital model} \label{gsmodel}
        One may also argue that using a synthetic population of NEOs can be useful to test and validate the methodology used in this work;
        such synthetic data do not contain any genetically related objects. The NEOSSat-1.0 orbital model (Greenstreet et al. 2012) is 
        widely regarded as the best model available to describe the orbital distribution of the NEO population. It is the result of 
        extensive integrations and therefore its results must reflect the effects of the web of overlapping resonances pointed out above. 
        Data sets generated in the framework of this model are free from the possible contamination of genetically related asteroids ---real 
        data are not, in principle. However, they are affected by the effects of secular resonances that permeate the entire NEO orbital
        domain; consequently this synthetic population may not be useful to further test our methodology.

\begin{figure}
        \centering
         \includegraphics[width=\linewidth]{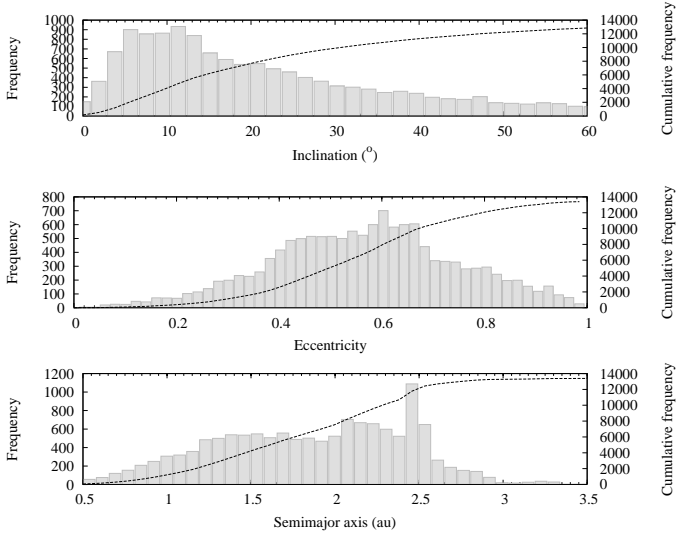}
         \caption{Distribution of the values of the orbital elements $a$, $e$, and $i$ for a synthetic sample of NEOs generated using the
                  NEOSSat-1.0 orbital model (see the text for details).
                 }
         \label{histosynthe}
      \end{figure}

        For this experiment, we have used the codes described in Greenstreet et al. (2012)\footnote{http://www.phas.ubc.ca/$\sim$sarahg/n1model/} 
        with the same standard input parameters to generate sets of orbital elements including 13\,392 virtual objects. Figure 
        \ref{histosynthe} shows a typical outcome. The mean values of $a$, $e$, and $i$ are 1.90 au, 0.58, and 23\fdg52, respectively; the 
        median values of $a$, $e$, and $i$ are 1.94 au, 0.58, and 18\fdg48, respectively. The respective IQR values are 0.91 au, 0.24, and 
        22\fdg45; the upper quartiles are 2.36 au, 0.69, and 32\fdg80, respectively. If we consider the values of the parameters associated 
        with the data in Fig. \ref{histo}, we observe that the two distributions are very different in terms of the values of the 
        inclination. There is also an excess of objects with values of the semimajor axis close to 2.5 au which corresponds to asteroids 
        escaping the main belt through the 3:1 mean-motion resonance with Jupiter; the Alinda family of asteroids (see e.g. Simonenko, 
        Sherbaum \& Kruchinenko 1979; Murray \& Fox 1984) are held by this resonance ---members of this family include 887 Alinda (1918 DB), 
        4179 Toutatis (1989 AC) and 6489 Golevka (1991 JX).   
%
%
%
%

\begin{figure}
       \centering
        \includegraphics[width=\linewidth]{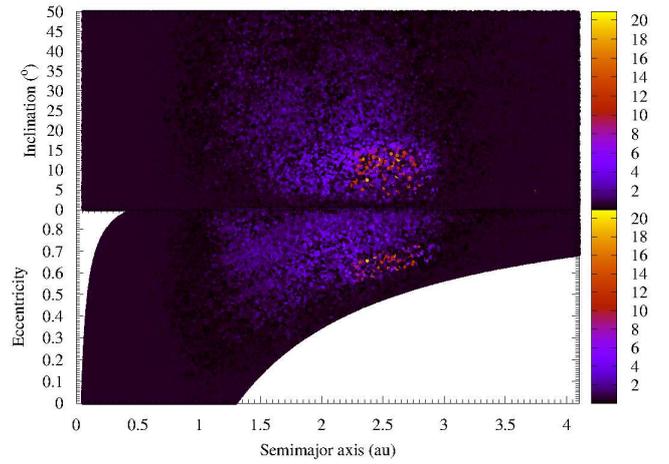}
        \caption{As Fig. \ref{maps}, right-hand panel but for synthetic data generated using the NEOSSat-1.0 orbital model (see Fig. 
                 \ref{histosynthe} and the text for details).
                }
        \label{mapgs}
     \end{figure}

        Figure \ref{mapgs} is equivalent to the right-hand panel in Fig. \ref{maps} but for the NEOSSat-1.0 orbital model. In this case,
        $\langle{c}\rangle=3.30\pm7.01$, and the cluster with the highest significance (20.6$\sigma$) includes 148 members and has $a=2.60$ 
        au, $e=0.63$, and $i=12\fdg6$. The Taurid Complex is missing in its entirety. In any case, if the Taurid Complex is the result of a
        cometary breakup, it cannot be reproduced by a model that neglects fragmentation. Unfortunately, this synthetic population of NEOs 
        produces a significant number of objects moving in Alinda-like orbits and perhaps may not be useful to provide a non-biased 
        benchmark to test the reliability of our significances; the two previous tests are much more informative and reliable in this case.
%
%
%
%

     \subsection{Changing the value of the cut-off parameter: effects on our results}
        In the previous sections, we have computed the $D$-criteria with all the NEOs currently catalogued and counted the number of minor 
        bodies with both $D_{\rm LS}$ and $D_{\rm R} < 0.05$. Therefore the value of the cut-off parameter for the countings performed has 
        been 0.05, but if we change the value of the cut-off parameter to e.g. 0.01 or 0.1, what are the effects on our results? Is 0.05 the 
        right choice for this study? 
%
%
     \begin{figure*}
       \centering
        \includegraphics[width=0.49\linewidth]{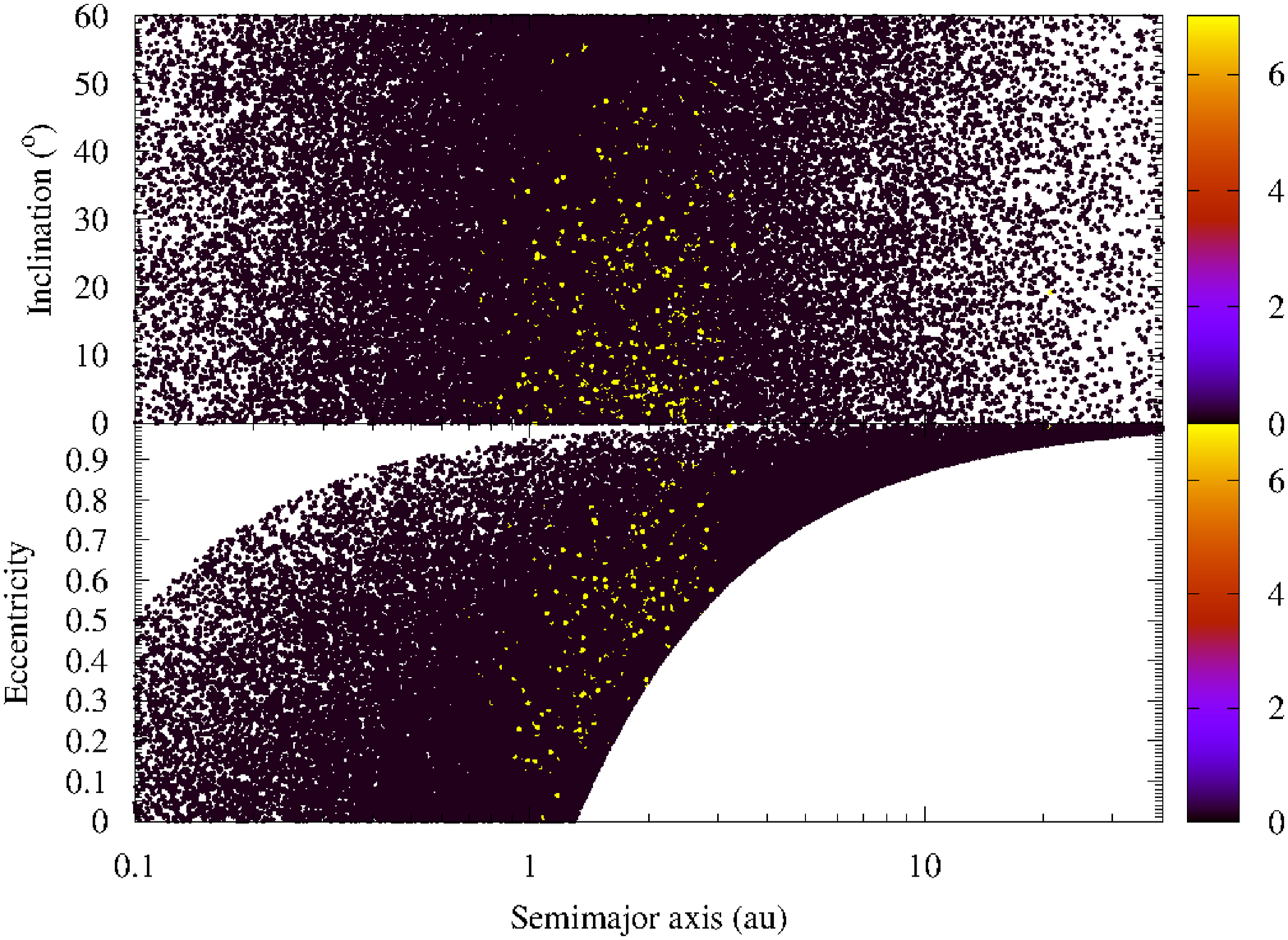}
        \includegraphics[width=0.49\linewidth]{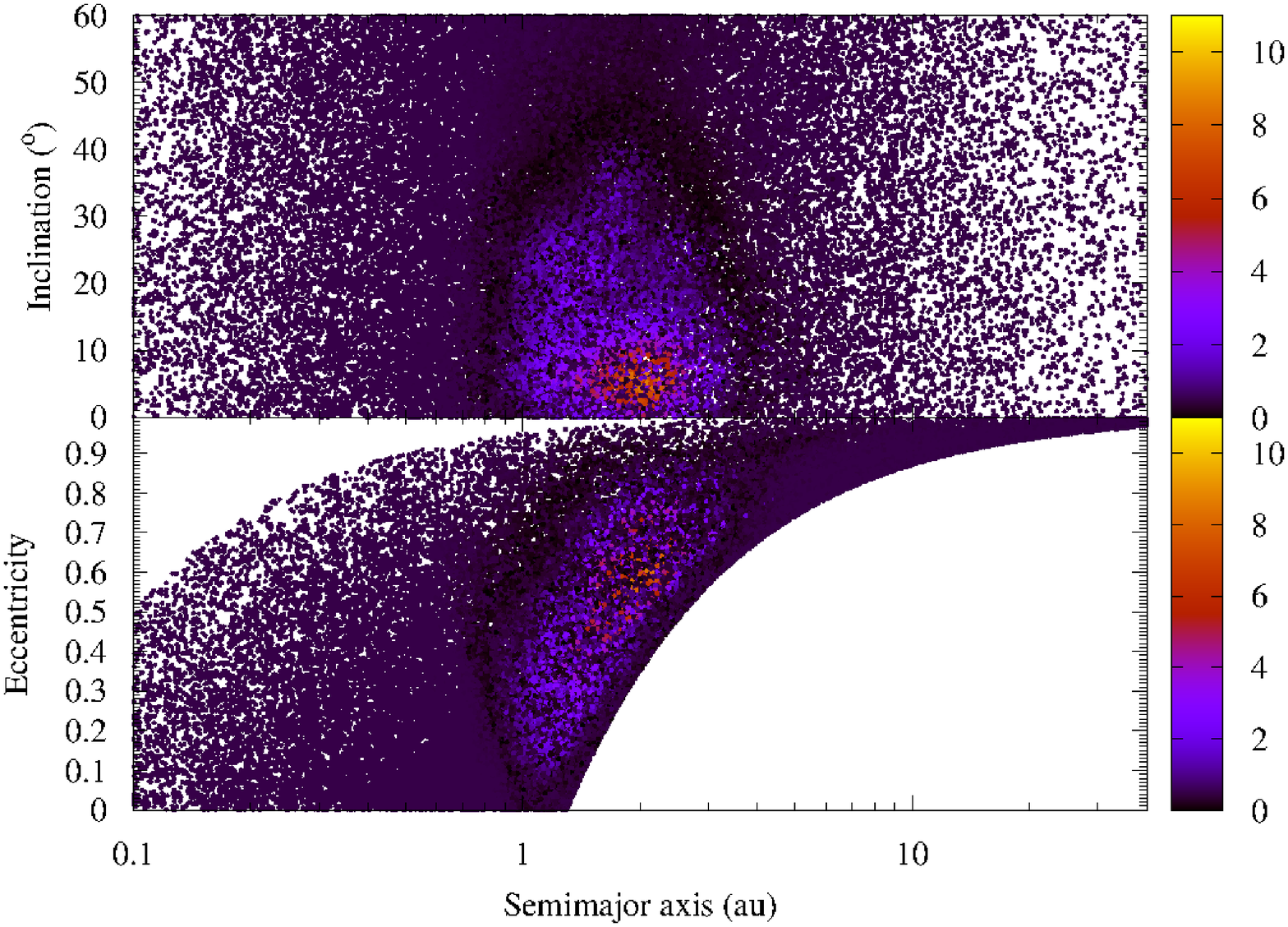}\\
        \caption{Equivalent to Fig. \ref{maps}, left-hand panel, but using $D_{\rm LS}$ and $D_{\rm R} < 0.01$ (left-hand panel) and $D_{\rm 
                 LS}$ and $D_{\rm R} < 0.1$ (right-hand panel).
                }
        \label{mapsx}
     \end{figure*}
%
%

        We have repeated the large simulation with 1$\times10^6$ sets of orbital elements but adopting a value of the cut-off parameter of 
        0.01 ---i.e. counted the number of minor bodies with both $D_{\rm LS}$ and $D_{\rm R} < 0.01$--- and the results are quite revealing 
        (see Fig. \ref{mapsx}, left-hand panel). First, given a random NEO, the probability of not having any other NEO within the volume of 
        the orbital parameter space defined by the value of the cut-off parameter is $>0.98$. Secondly, the largest dynamical groupings have 
        just three members and, more importantly, there are no groupings linked to comet 73P/Schwassmann-Wachmann 3, the only bona fide 
        (genetic) asteroid cluster. These results are clearly indicating that a value of the cut-off parameter of 0.01 is too restrictive 
        and physically unjustified.

        Increasing the value of the cut-off parameter to perform the countings ($D_{\rm LS}$ and $D_{\rm R} < 0.1$) has also dramatic 
        effects (see Fig. \ref{mapsx}, right-hand panel). Now the probability of not having any dynamically related NEO drops to 0.41 and 
        the largest grouping has 327 members. The largest dynamical grouping is found for $a=2.0335$ au, $e=0.5950$, and $i=5\fdg12583$ 
        (perhaps linked to the Taurid Complex). Because $\langle{c}\rangle=16.9\pm33.8$, the statistical significance of this grouping (as 
        defined above) is over 9$\sigma$. In sharp contrast, the dynamical groupings linked to comet 73P/Schwassmann-Wachmann 3 have now a 
        significance $<1.8\sigma$. Therefore, a value of the cut-off parameter of 0.1 is too large and unsuitable to recover true dynamical 
        groupings in a reliable manner.

        Our analysis of the effects of the value of the cut-off parameter on our results confirms that, as pointed out in Rudawska et al. 
        (2012), 0.05 is physically justified and it is the right choice for this study. 

     \subsection{Subtracting the background: yet another assessment of the statistical significance}
\begin{figure*}
       \centering
        \includegraphics[width=0.49\linewidth]{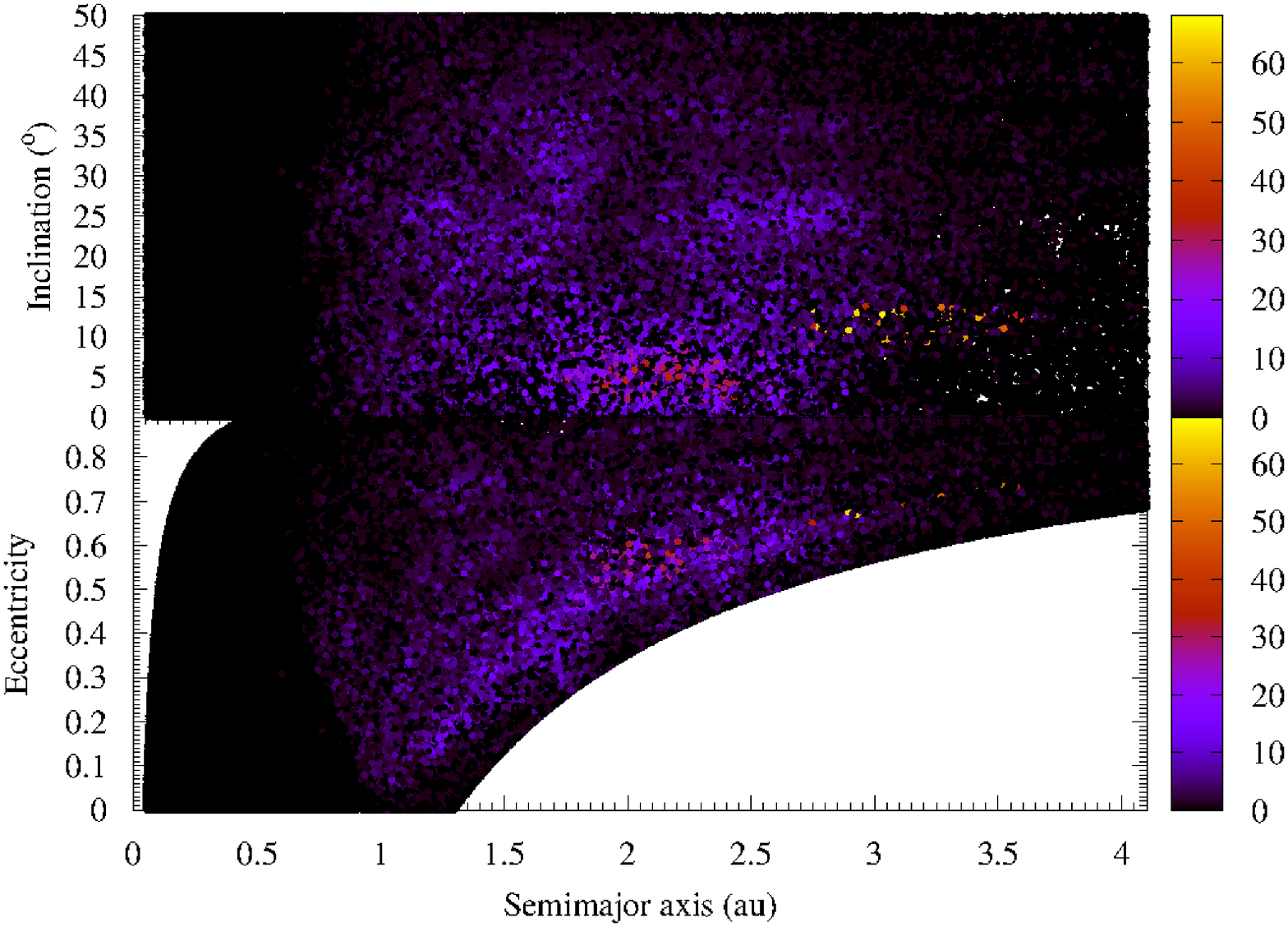}
        \includegraphics[width=0.49\linewidth]{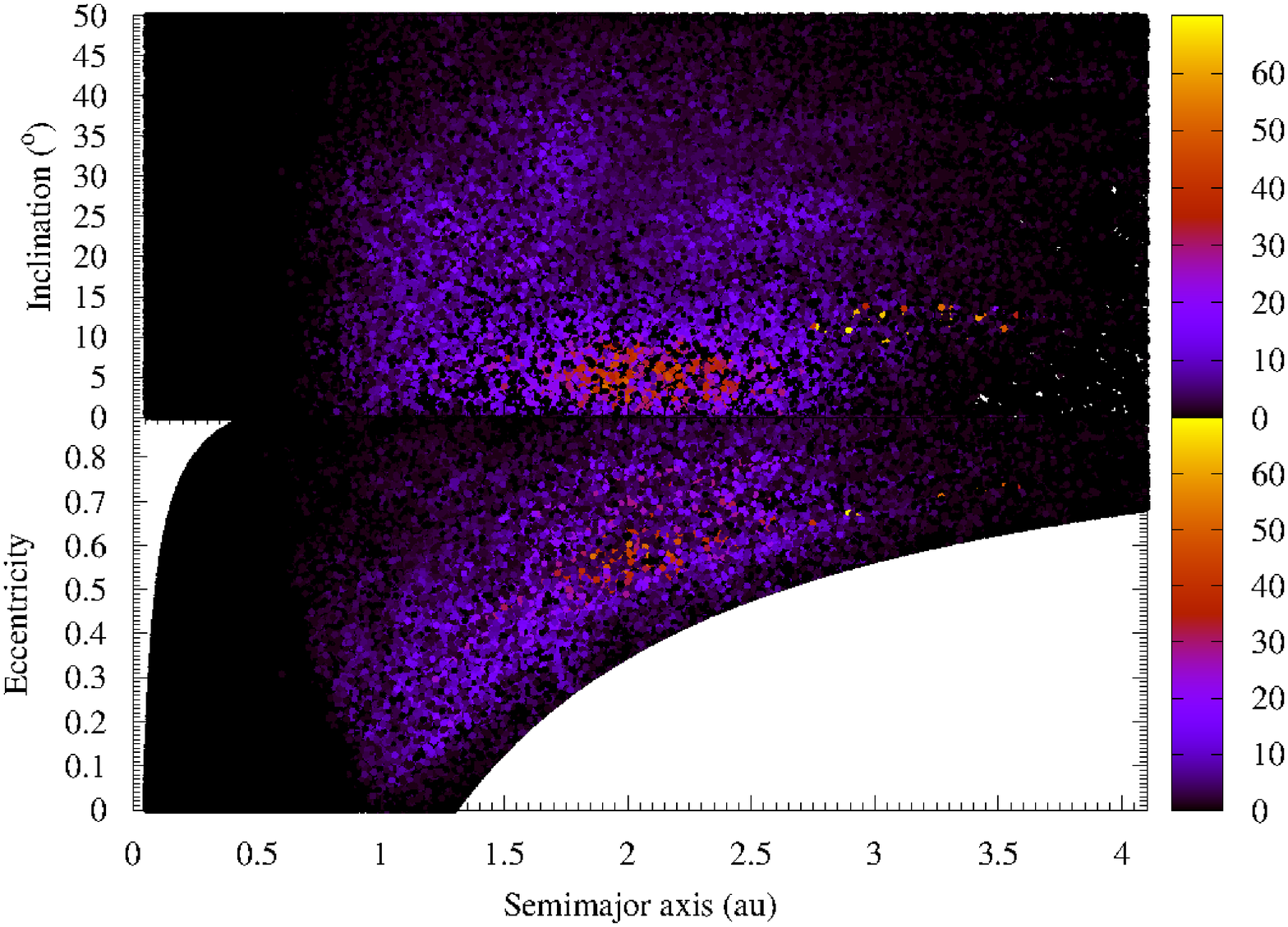}\\
        \includegraphics[width=0.49\linewidth]{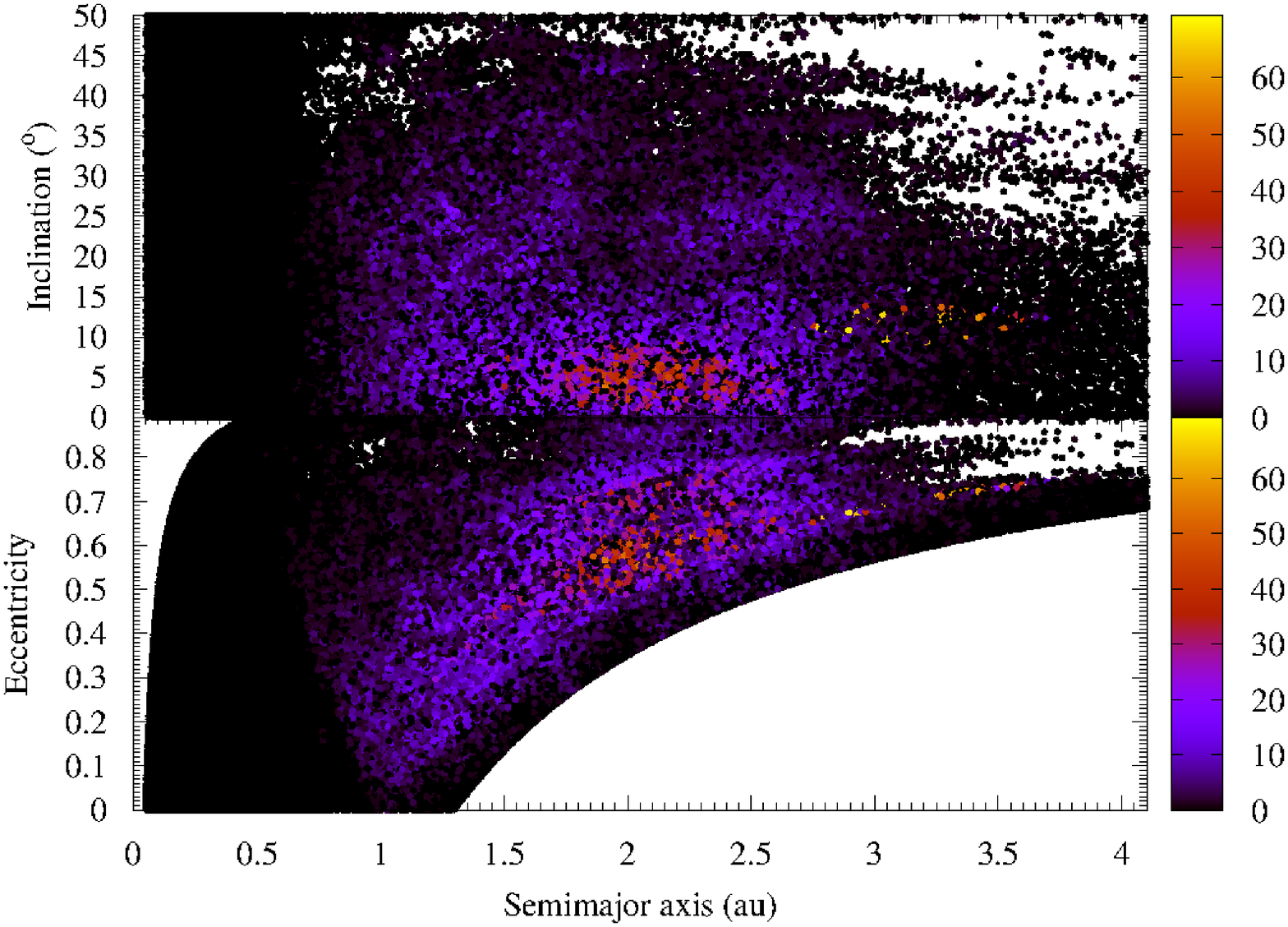}
        \includegraphics[width=0.49\linewidth]{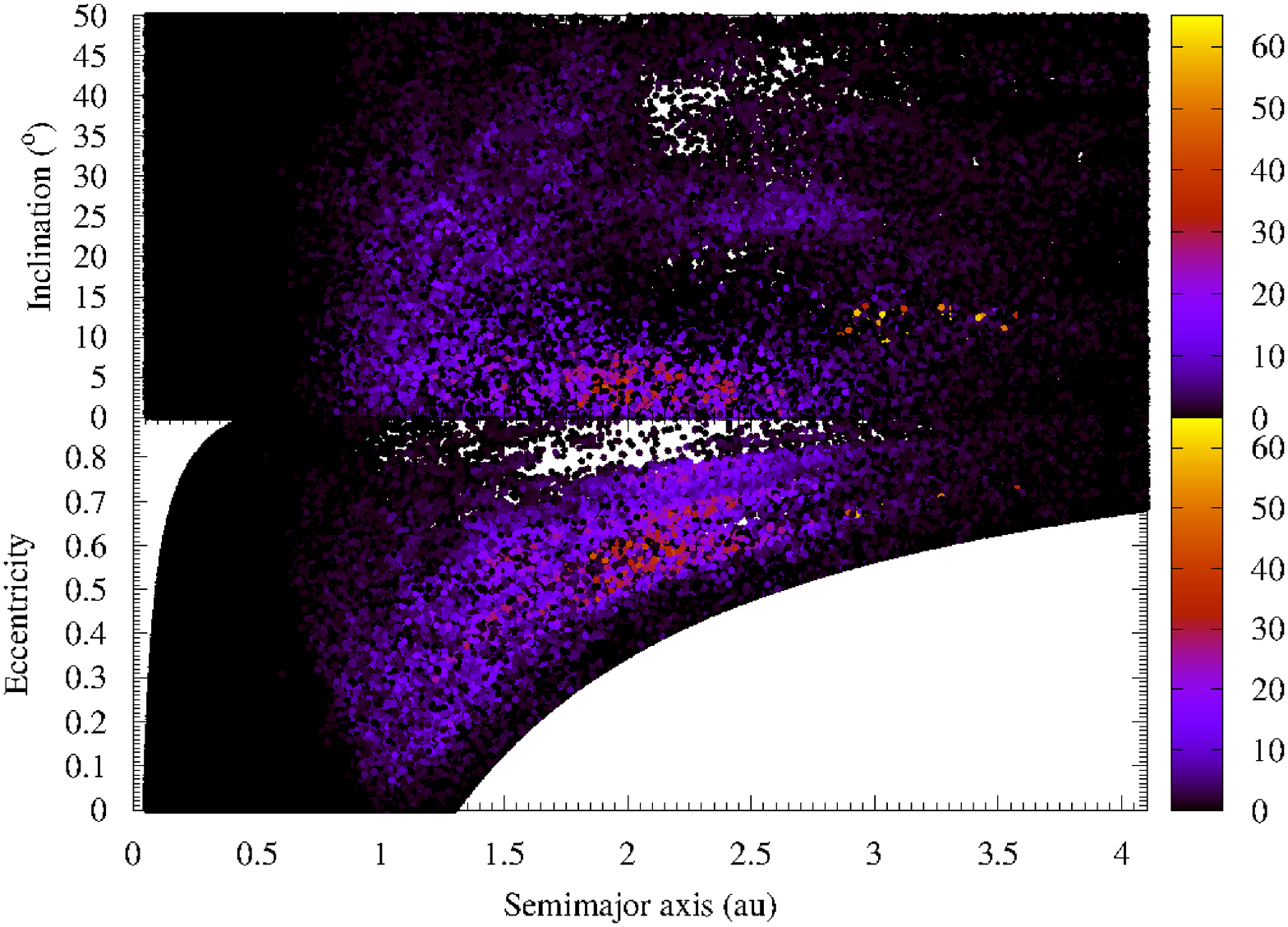}\\
        \caption{Subtracting the background. In these colour maps the number of counts ($D_{\rm LS}$ and $D_{\rm R} < 0.05$) per virtual 
                 NEO minus that of relevant reference samples is plotted: counts after subtracting altered data with Gaussian noise at the 
                 30 per cent level (left-hand-top panel), counts after subtracting altered data with Gaussian noise at the 90 per cent level 
                 (right-hand-top panel), counts after subtracting a putative uniform background (left-hand-bottom panel), and actual NEOs
                 minus the NEOSSat-1.0 orbital model (right-hand-bottom panel).  
                }
        \label{subs}
     \end{figure*}

        In Fig. \ref{maps}, right-hand panel, Figs \ref{map30} and \ref{map90}, and Fig. \ref{mapgs} we have used the same random mesh to 
        compute the $D$-criteria and perform the countings. This lets us subtract counts easily to compare results from real and synthetic
        data, further assessing the statistical significance of our dynamical groupings. In the colour maps in Fig. \ref{subs} we have 
        subtracted the counts from the various synthetic populations of NEOs discussed above from the original one. For the top panels the 
        data from Fig. \ref{maps}, right-hand panel, and those in Figs \ref{map30} and \ref{map90} have been used. For the left-hand-bottom 
        panel, data from Fig. \ref{maps}, right-hand panel, and from Fig. \ref{mapu} have been used. Finally, for the right-hand-bottom 
        panel, data from Fig.  \ref{maps}, right-hand panel, and from Fig. \ref{mapgs} have been used.
%
%
%
%

        Figure \ref{subs}, right-hand-bottom panel, shows the difference in counts between the sample used here and an instance of the
        NEOSSat-1.0 orbital model (data in Section \ref{gsmodel}). This subtracted plot is used here for qualitative understanding of the
        features in Fig. \ref{best} rather than for a quantitative comparison with the data, which is beyond the scope of this paper. As
        expected, the NEOSSat-1.0 orbital model does not account for the comet 73P/Schwassmann-Wachmann 3 group. The substructure seen in
        Fig. \ref{best} stands out clearly. Results from the NEOSSat-1.0 orbital model are the outcome of extensive numerical integrations
        assuming certain sources for the NEOs (for details, see Greenstreet et al. 2012). Therefore, some of the features in Fig. \ref{best}
        must have a different origin. The NEOSSat-1.0 orbital model does not include fragmentation and strictly applies only if $H<22$~mag
        (originally $H<18$~mag). About 47.8 per cent of NEOs in our sample have $H>22$~mag. Many of these objects could be fragments or even 
        fragments of fragments, therefore the comparison is not very realistic and any conclusions obtained are weakened by these facts. The 
        other three panels in Fig. \ref{subs} confirm that the significance maps in Fig. \ref{maps} are robust and reliable because this 
        time the colour maps do not rely on global information like $\langle{c}\rangle$ and $\sigma_{\rm c}$ but on point-to-point 
        subtraction of counts.

\begin{figure}
        \centering
         \includegraphics[width=\linewidth]{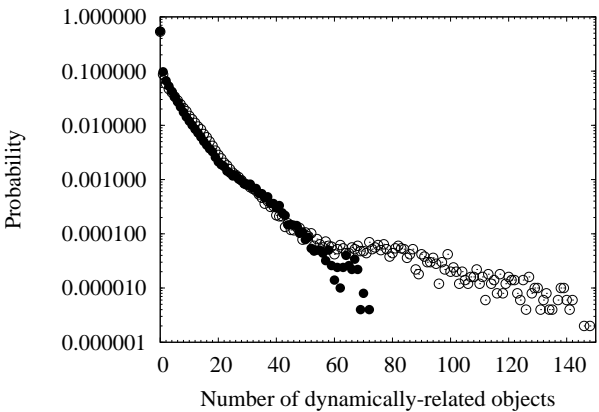}
         \caption{Probability of having $X$ dynamically related objects in the experiments plotted in Fig. \ref{maps}, right-hand panel 
                  (filled circles), and Fig. \ref{mapgs} (empty circles).  
                 }
         \label{probC}
      \end{figure}

        Figure \ref{probC} shows the probability of having $X$ dynamically related objects in the experiments plotted in Fig. \ref{maps}, 
        right-hand panel, and Fig. \ref{mapgs}. The respective probabilities of having no dynamically related objects, $P(0)$, are 0.54249
        and 0.533358. There is an obvious similarity between the two distributions, but only for $X<50$--60. The lack of similarity beyond 
        that point could be the result of statistical fluctuations or perhaps indicate that fragmentation issues could be important for 
        large dynamical groupings if they do exist.   
%
%
%
%

  \section{Relevant dynamical groupings}
     In this section, we further explore some relevant and distinct clusterings but given the very complex substructure present in Figs 
     \ref{best} and \ref{soso}, we do not attempt an exhaustive study. Our findings are consistent with an analysis that suggests that the 
     group of recent, most powerful Earth impacts do not follow a strictly random pattern (de la Fuente Marcos \& de la Fuente Marcos 2015b).

     \subsection{The Taurid Complex}
        Although the Taurid Complex is the best-studied structure present in Fig. \ref{maps}, Fig. \ref{best} shows that it appears to make 
        a relatively minor contribution to the overall NEO dynamical substructure. Assuming the boundaries discussed in Steel \& Asher 
        (1996), e.g. $1.8<a<2.6$~au, $0.64<e<0.85$, and $i\leq12$\degr, the Taurid Complex is somewhat confined between the curves with
        perihelion at 0.458 au and 0.65 au in Fig. \ref{best}. Discarding outliers, clustering associated with the Taurid Complex tends to
        exhibit the highest statistical significance (the most significant cluster, 11.7$\sigma$, is at $a\sim2$~au, $e\sim0.6$, and 
        $i\sim4$\degr and includes 60 members). The eccentricities of these Taurid Complex asteroids are however lower than those of the 
        Taurid meteor streams described in Porub\v{c}an, Korno\v{s} \& Williams (2006).

     \subsection{The Comet 73P/Schwassmann-Wachmann 3 group}
        The grouping with the highest statistical significance ($\sim$13.5$\sigma$) and the lowest values of the $D$-criteria is linked to 
        comet 73P/Schwassmann-Wachmann 3 ($a$=3.06~au, $e$=0.70, and $i$=11.38\degr). This comet started to disintegrate in 1995 and 
        multiple fragments were observed in 2006 and 2007 (Crovisier et al. 1996; Weaver et al. 2006; Reach et al. 2009); it may consist of 
        hundreds of fragments now (66 are included in SBDB). This disruption event is responsible for the outliers in Fig. \ref{all} and 
        elsewhere. Groupings including these objects organize themselves between the curves associated with $q=0.9$ au and $q=0.983$ au in 
        Fig. \ref{best}. These groupings contain a fraction of interlopers as high as 10 per cent and include as many as 72 members. 

     \subsection{The 5011 Ptah (6743 P-L) group}
        NEOs with $a\in(1.48, 1.9)$~au, $e\in(0.5, 0.6)$, and $i\in(2, 8)$\degr define multiple statistically significant groupings at the 
        5$\sigma$ level or better (see Fig. \ref{best}). This orbital parameter subdomain contains 180 NEOs, the only named object in the 
        list is 5011 Ptah (6743 P-L) although it is not the largest of the group ($H$ = 16.4 mag), 86039 (1999 NC$_{43}$) is probably larger 
        ($H$ = 16.0 mag); the smallest member of the group is 2002~SQ$_{222}$ ($H$ = 30.1 mag). Two objects from this group, 86039 
        (Borovi\v{c}ka et al. 2013) and 2011~EO$_{40}$ (de la Fuente Marcos \& de la Fuente Marcos 2013, 2014; de la Fuente Marcos, de la 
        Fuente Marcos \& Aarseth 2015), have been linked to the Chelyabinsk impactor. Reddy et al. (2015) have pointed out that the 
        existence of a connection between the Chelyabinsk meteoroid and 86039 is rather weak, both in dynamical and compositional terms. 
        Besides there is yet no spectroscopic evidence linking 2011~EO$_{40}$ to Chelyabinsk. However, the orbit of the actual impactor was 
        almost certainly part of this dynamical grouping (see the extensive discussion in de la Fuente Marcos et al. 2015). NEOs in this 
        group reach perihelion in the region between 0.65 au and 0.82 au that is controlled by the Kozai mechanism.

     \subsection{The 85585 Mjolnir (1998 FG$_{2}$) group}
        The orbital parameter subdomain loosely defined by $a\in(1.28, 1.42)$~au, $e\in(0.3, 0.4)$, and $i\in(2, 7)$\degr includes several 
        statistically significant groupings at the 5$\sigma$ level (see Fig. \ref{best}), this represents about 30 objects matching the 
        criterion presented in Section 2 when the expected number at the 3$\sigma$ level is about half that figure. In total, there are 73 
        objects within that volume of the orbital parameter space, the only named object in the list is 85585 Mjolnir (1998 FG$_{2}$) 
        although it is not the largest of the group ($H$ = 21.6 mag), nine others are larger; the smallest members of the group are 2008 VM 
        ($H$ = 30.2 mag) and 2008~TC$_{3}$ ($H$ = 30.4 mag). Meteoroid 2008~TC$_{3}$ caused the Almahata Sitta event (Jenniskens et al. 
        2009; Oszkiewicz et al. 2012). Mjolnir has an F-class spectrum similar to that of 2008~TC$_{3}$ (Jenniskens et al. 2010), their 
        orbits are akin. Objects in this dynamical group reach perihelion at 0.82--0.9~au in the region dominated by secular resonances.

     \subsection{The 101955 Bennu (1999 RQ$_{36}$) group}
        The orbital parameter subdomain loosely defined by $a\in(1.10, 1.17)$~au, $e\in(0.18, 0.25)$, and $i\leq10$\degr includes some
        marginally significant groupings at the 3--5$\sigma$ level (see Fig. \ref{soso}). There are 43 objects within that volume of the 
        orbital parameter space, the only named object in the list is 101955 Bennu (1999 RQ$_{36}$) although it is not the largest of the 
        group ($H$ = 20.9 mag), two others are larger; the smallest member of the group is 2014~AA ($H$ = 30.9 mag). Meteoroid 2014~AA 
        collided with our planet on 2014 January 2 (Chesley et al. 2014, 2015; Kowalski et al. 2014). Objects in this group also reach 
        perihelion at $\sim$0.9~au.

  \section{Discussion}
     Our counting experiments above are characterized by values of the standard deviations which are larger than their respective means.
     This is not at all surprising, even Fig. \ref{histo} clearly shows that ---within the NEO orbital parameter space--- there are 
     preferred places and avoided places. In addition, outliers increase the value of the standard deviation. Assuming that the observed 
     distribution in orbital parameter space is not fully shaped by observational bias, it is unlikely to find many objects following 
     unstable orbits simply because they remain there for a short time. The probabilities of finding an asteroid are not similar and 
     constant; there are regions of the orbital parameter space where asteroids are rare and others where asteroids are frequent. This
     automatically induces overdispersion, i.e. the variance is larger than the mean. 

\begin{figure}
        \centering
         \includegraphics[width=\linewidth]{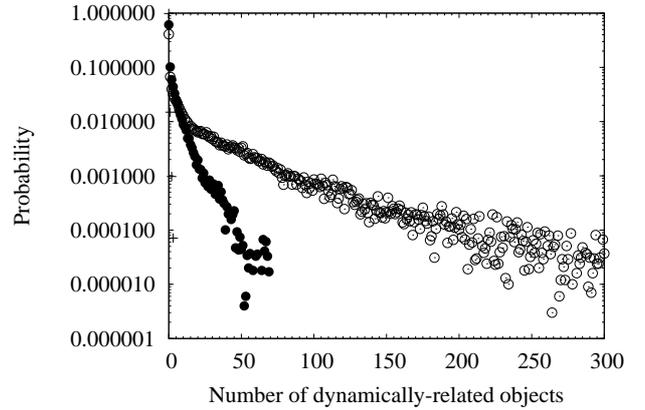}
         \caption{Probability of having $X$ dynamically related objects in the experiments plotted in Fig. \ref{maps}, left-hand panel, and
                  Fig. \ref{mapsx}. Crosses signal the probabilities for a value of the cut-off parameter of 0.01, filled circles for 0.05, 
                  and empty circles for 0.1.
                 }
         \label{prob}
      \end{figure}

     Focusing on dynamical groupings with $D_{\rm LS}$ and $D_{\rm R} < 0.05$, the observed overdispersion is due to two factors. First,
     the probability of having a similar orbit is not the same for all pairs of NEOs. Secondly, not all the objects have the same tendency
     to have others nearby in terms of the metrics used here. NEOs captured in stable resonances are more likely to have others in their
     dynamical neighbourhood. On the other hand, NEOs resulting from breakups of any kind are also more likely to have other NEOs 
     following similar orbits. Such objects are less likely to have a few (or none) dynamically related companions. We have found a large 
     variation in individual tendencies to be part of a dynamical grouping. This degree of overdispersion is mostly induced by excessive
     zero counts that lead to underestimate the variance of the number of dynamically related NEOs which in turn overstates the 
     significance of the dynamical groupings. However, this is not a concern in our case because when we changed the value of the cut-off 
     parameter to 0.1, we lowered the probability of having exactly zero or one dynamically related object and the statistical significance 
     of most relevant dynamical groupings remained nearly the same (compare Fig. \ref{maps}, left-hand panel, with Fig. \ref{mapsx}, 
     right-hand panel but an obvious exception was the comet 73P/Schwassmann-Wachmann 3 group). Figure \ref{prob} shows the probability of 
     having exactly 0, 1,... objects for a value of the cut-off parameter of 0.01 (crosses), 0.05 (filled circles), and 0.1 (empty circles). 
     The respective probabilities of having no dynamically related objects, $P(0)$, are 0.984026, 0.61017, and 0.409548. For a value of the 
     cut-off parameter of 0.01, $P(>3)=0$; for 0.1, $P(\geq3)=0.484$ 
%
%
%
%

     It may be argued that the substructure present in Fig. \ref{best} is not more than a reflection of the distributions of the orbital 
     elements of the NEO population showed in Fig. \ref{histo} but this is clearly incorrect as many of the statistically significant 
     groupings have values of the orbital parameters far from the most probable ones (median values), being part of the first or fourth 
     quartiles of the distribution (see above). The substructure uncovered here must be real; our methodology is sensitive enough to 
     identify a well-documented disruption event, that of comet 73P. On the other hand, the clustering discussed here is not like that of 
     the classical asteroid families (see e.g. the review in Nesvorn\'y et al. 2015) found in the main asteroid belt. The NEO dynamical 
     groupings have higher eccentricities and represent excesses in orbital parameter space with respect to some arbitrarily defined, but 
     nonetheless statistically reasonable, background level (see Section 4).

     Schunov\'a et al. (2012) showed that, using the data available at that time, it was not possible to find a statistically significant
     asteroid cluster with four or more members and one may be tempted to conclude that our results are dramatically different when compared 
     to theirs. This conclusion is incorrect and we have to emphasize that even if our work is somewhat connected with that in Schunov\'a et 
     al. (2012), our goals are intrinsically different. Schunov\'a et al. (2012) focus on searching for genetically related asteroids among 
     the NEO population and use the $D_{\rm SH}$ criterion to that end. In sharp contrast, our work studies statistically significant 
     dynamical groupings that may or (more likely) may not have any physical connection; we do not use $D_{\rm SH}$ but $D_{\rm LS}$ and 
     $D_{\rm R}$. As pointed out above, our approach can detect genetically related asteroid clusters (see Section 5.2), but this is not its 
     primary objective. Our results are fully consistent with those in Schunov\'a et al. (2012) if we use $D_{\rm SH}$, with the sole 
     exception of the genetically related clusters associated with comet 73P as Schunov\'a et al. (2012) excluded comets from their 
     analysis.

\begin{figure}
        \centering
         \includegraphics[width=\linewidth]{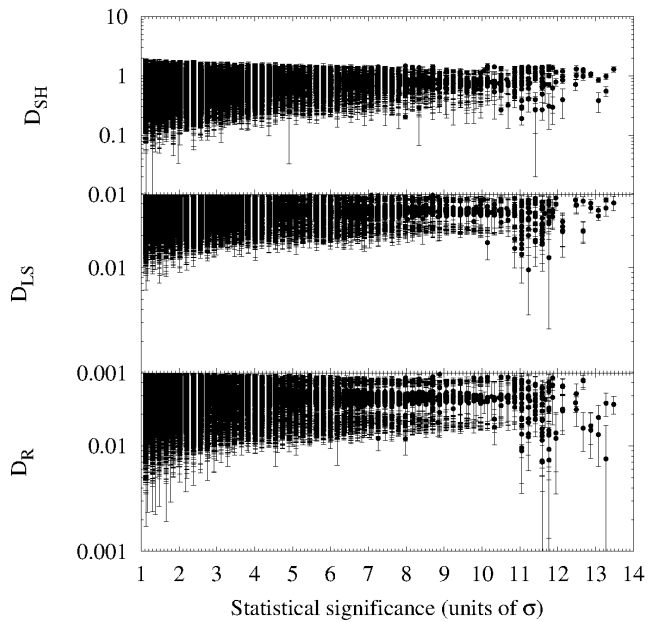}
         \caption{Average values of the various $D$-criteria as a function of the statistical significance for the data in Fig. \ref{all}.
                  Error bars give one standard deviation.
                 }
         \label{Ds}
      \end{figure}

     As an additional proof that our results are not at odds with those in Schunov\'a et al. (2012), we have represented the average values 
     of $D_{\rm SH}$, $D_{\rm LS}$, and $D_{\rm R}$ as a function of the statistical significance (see Fig. \ref{Ds}) for all the groupings 
     in Fig. \ref{all}. In the figure, the error bars represent the standard deviation. The top panel in Fig. \ref{Ds} clearly shows that the 
     groupings with the lowest average values of $D_{\rm SH}$ (those more likely to include asteroid siblings) have statistical significance 
     under 3$\sigma$ which is fully consistent with the conclusions in Schunov\'a et al. (2012) even if the input data sets are different. 
     One may also wonder if the groupings with the lowest average values of $D_{\rm SH}$ in Fig. \ref{Ds} are somewhat linked to the objects 
     of interest discussed in Schunov\'a et al. (2012). These groupings have between 4 and 8 members, their $D_{\rm SH}$ is $\sim0.08$, and 
     $a\in(0.98, 1.04)$ au, $e\in(0.02, 0.08)$, and $i<8$\degr. These values are more or less consistent with those of the C3 NEO cluster in 
     Schunov\'a et al. (2012) and correspond to objects moving in Earth-like orbits, the Arjuna asteroids (Cowen 1993; Rabinowitz et al. 
     1993; Gladman, Michel \& Froeschl\'e 2000). This group of peculiar objects has been recently reviewed in detail by de la Fuente Marcos 
     \& de la Fuente Marcos (2015a) and they are unlikely to form a genetically related family although asteroids 2014~EK$_{24}$ and 2013 
     RZ$_{53}$ (de la Fuente Marcos \& de la Fuente Marcos 2015c) follow very similar orbits (including five orbital elements). Objects in 
     this group are trapped in a web of resonances that keep many of them within the Earth's co-orbital zone for tens of thousands of years 
     as they experience repeated resonant episodes of the horseshoe, quasi-satellite, or Trojan type.  
%
%
%
%

     On the other hand, the groupings with the highest significance in Fig. \ref{Ds} have relatively high values of $D_{\rm LS}$, and 
     $D_{\rm R}$ because they include fractions of interlopers of 5 per cent or higher. This further supports the conclusion that most of
     the observed dynamical substructure has its roots in secular resonances not in breakup events. Without a doubt, catastrophic failure
     of NEOs happened in the past history of the Solar system (i.e. the Taurid Complex), but it is also happening now (see the case of comet 
     73P above) and it will continue taking place in the future. However, these events seem to provide just a small contribution to the 
     present-day, overall amount of orbital coherence observed within the NEO population although this contribution is in the form of the
     groupings with the highest statistical significance. In general, the results of our analysis are more consistent with secular 
     resonances and the Kozai mechanism being the sources of most of the observed dynamical coherence; NEOs are being confined via 
     resonances into specific paths with well-defined perihelia. It is now widely accepted that groups of objects moving initially in 
     similar trajectories (debris resulting from breakups) lose all orbital coherence in a short time-scale (Pauls \& Gladman 2005; Rubin 
     \& Matson 2008; Lai et al. 2014). In contrast, objects confined by resonances can remain together, dynamically speaking, for longer 
     time-scales and those that are lost can be easily replaced via resonant capture. Long-term stable asteroid streams could be real but 
     not linked to discrete breakups but to the pervasive architecture of the NEO orbital realm.

     The topic of impacts linked to groups of asteroids has been discussed in the past. Halliday, Blackwell \& Griffin (1990) provided early 
     evidence for groups of meteor events and, within these groups, preference for perihelia just slightly inside the Earth's orbit. Data in 
     their table 3 suggests that small impactors with values of $a$ in the range 1.7--2.5~au, $e$ in 0.4--0.6, and $i$ in 0--12\degr are 
     more likely. Based on Halliday et al. (1990) results, Drummond (1991) found evidence for asteroid streams linked to the groups of 
     meteor events. However, a more detailed study by Fu et al. (2005) concluded that it is unlikely that the streams found in Drummond 
     (1991) be anything more than random fluctuations in the orbital parameter space crossed by the NEO population. Benoit \& Sears (1995) 
     used data on modern falls of chondrites and found that most meteorite parent bodies had $q\sim1$~au, with only a small fraction 
     ($\sim14$ per cent) having orbits with $q<0.85$~au. They observed a tendency for meteorites with large cosmic ray exposure ages ($>35$ 
     Myr) to have shorter perihelia, $q\sim0.8$~au, than those with relatively short cosmic ray exposure ages, which tend to have perihelia 
     between 0.85~au and 1.0~au. This may suggest that younger material tends to drift their perihelia towards the 0.85--1.0~au region, 
     perhaps as a result of non-gravitational forces. Alternatively, one may assume that smaller fragments should be, in general, younger 
     than larger bodies and the groupings identified here include both large and very small bodies. Large NEO fragments (size between 200~m
     and 700~m) are unlikely to have been produced during the last few thousand years but meteoroids may have been ejected via e.g. 
     rotational or tidal instability after the parent body has already been trapped within a web of secular resonances. In this scenario, 
     NEOs with perihelia closer to the orbit of the Earth are more likely to produce fragments which is a reasonable hypothesis. 

     Figure \ref{mapsx}, left-hand panel, in which the groups with the highest significance have two-to-three members links directly to the 
     topic of unbound asteroid pairs (see e.g Vokrouhlick\'y \& Nesvorn\'y 2008; Pravec \& Vokrouhlick\'y 2009; Pravec et al. 2010; 
     Moskovitz 2012; Duddy et al. 2013; Polishook et al. 2014; Wolters et al. 2014). Bona fide unbound asteroid pairs have been found in the 
     main asteroid belt. The type of analysis performed in Fig. \ref{mapsx} signals the presence of multiple candidate pairs among the NEO 
     population, although many of them could be just coincidental couples from the background asteroid population.

  \section{Conclusions}
     In this paper, we have shown statistically that the distribution of the NEO population in orbital parameter space is far from random.
     We have confirmed the presence of statistically significant dynamical groupings among the NEO population that seem to be associated 
     with the secular resonant architecture described in Michel \& Froeschl\'e (1997). Some of these dynamical groupings appear to have 
     been the immediate sources of recent asteroid impact events and are likely to continue doing so in the future. The groupings host both
     relatively large (size between 200~m and 700~m) and very small (a few metres) objects which suggest that some of the smaller fragments 
     may have been produced {\it in situ} via rotational instability or other mechanisms (see e.g. Denneau et al. 2015). Production of 
     meteoroids within the immediate neighbourhood of the Earth has a direct impact on the evaluation of the overall asteroid impact hazard 
     (e.g. Schunov\'a et al. 2014). However and although this is of considerable theoretical interest, most of these fragments are small 
     enough to be of less concern in practice. 

  \section*{Acknowledgements}
     The authors thank the referee, M. Granvik, for his constructive, detailed and very helpful reports. In preparation of this paper, we 
     made use of the NASA Astrophysics Data System and the ASTRO-PH e-print server.

  \bsp
  \label{lastpage}
\end{document}